# Repulsive photons in a quantum nonlinear medium

Sergio H. Cantu,[1, *] Aditya V. Venkatramani,[2, *] Wenchao Xu,[1] Leo Zhou,[2] Brana Jelenković,[3] Mikhail D. Lukin,[2] and Vladan Vuletić[1]

[1]Department of Physics and Research Laboratory of Electronics,
Massachusetts Institute of Technology, Cambridge, Massachusetts 02139, USA
[2]Department of Physics, Harvard University, Cambridge, Massachusetts 02138, USA
[3]Photonics Center, Institute of Physics, University of Belgrade, Serbia

**The ability to control strongly interacting light quanta (photons) is of central importance in quantum science and engineering [1–5]. Recently it was shown that such strong interactions can be engineered in specially prepared quantum optical systems [6–10]. Here, we demonstrate a method for coherent control of strongly interacting photons, extending quantum nonlinear optics into the domain of repulsive photons. This is achieved by coherently coupling photons to several atomic states, including strongly interacting Rydberg levels in a cold Rubidium gas. Using this approach we demonstrate both repulsive and attractive interactions between individual photons and characterize them by the measured two- and three-photon correlation functions. For the repulsive case, we demonstrate signatures of interference and self ordering from three-photon measurements. These observations open a route to study strongly interacting dissipative systems and quantum matter composed of light such as a crystal of individual photons [11, 12].**

Strong interactions between individual photons can be realized by coherently coupling them to strongly interacting Rydberg states inside an atomic gas using electromagnetically induced transparency (EIT) [13, 14]. Inside this optical medium, the photons travel as coupled excitations of light and matter called dark-state polaritons, and inherit interactions from their atomic components [15, 16]. When the medium is optically dense, the polariton propagates as a massive particle with a velocity that is much smaller than the speed of light, and interacts with other polaritons via the Rydberg atomic component [17]. This approach has been used to create absorptive non-linearities [8, 18], attractive interactions, resulting in the formation of two- and three-photon bound states [14, 19], and transistors [20, 21] at the single photon level. At the same time, realization of strong repulsive interactions, that are of interest for many potential applications in quantum metrology, quantum information, and quantum simulation of certain Hamiltonians [11, 12, 22], have proved much more challenging [23].

In this Letter we demonstrate a novel method for coherent control of strongly interacting photons, extending quantum nonlinear optics into the domain of repulsive photons. In contrast to previous Rydberg-EIT schemes, we make use of a four-level atomic level configuration (Fig.1A-1B) that couples probe photons to two distinct long-lived atomic states via two distinct two-photon processes. One of these long-lived states is a Rydberg state $|r\rangle$ that enables interactions between photons. The other is a long-lived hyperfine state $|f\rangle$ of the ground state manifold that provides additional control over the particle's group velocity and effective mass [24, 25]. A trans-mission spectrum for the probe photons in this scheme (Fig.1C) shows two nearly degenerate EIT resonances, corresponding to the two-photon resonances with states $|r\rangle$ and $|f\rangle$. Such a double-EIT scheme results in the formation of two dark-state branches $|d_0\rangle$ and $|d_1\rangle$ with strongly modified dispersion, as can be seen in Fig.1D. These dark-state branches combine both the interacting $|r\rangle$ and the non-interacting $|f\rangle$ states allowing us to control the effective interactions. The inverse of the slope and the curvature of the dispersion relation determine the group velocities $(v_{d0}, v_{d1})$ and masses $(m_{d0}, m_{d1})$ of these two dark-states, respectively. The propagation dynamics and non-linear interactions arising from Rydberg components, further result in a strong mixing between polaritons $|d_0\rangle$ and $|d_1\rangle$. This coupling between the two polariton states and the resulting modified dispersion provide the necessary flexibility to control separately the effective potential between photons and their effective mass. The inset to Fig.1D shows a case where increasing the coupling $\Omega_f$ to $|f\rangle$ switches the effective potential between photons from negative to positive, while their effective mass remains positive. Such tunability allows us to generate both attractive and repulsive interactions.

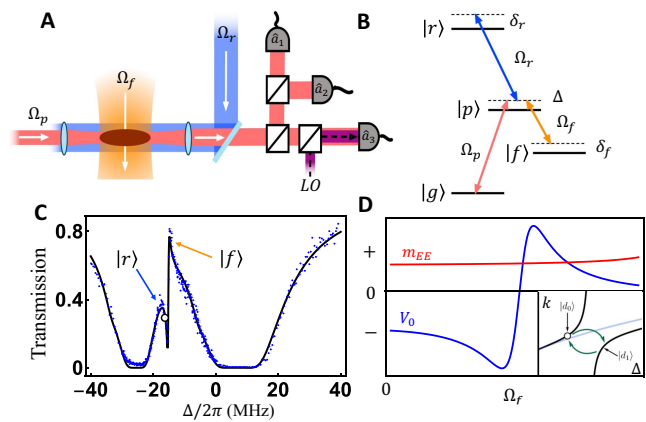

Figure 1: **Photon interaction engineering. A,B** Experimental setup and level structure. A weak probe field (red) is coupled to the Rydberg state $|r\rangle$ by a counter-propagating control field (blue) and to a non-interacting hyperfine ground state $|f\rangle$ by a secondary control field (orange) applied perpendicularly. Probe photons are split onto three single-photon detectors for correlation measurements. A local oscillator can be superimposed onto one of the detector paths to perform conditional phase measurements. **C** Transmission spectrum of the probe. The two peaks correspond to the two-photon EIT resonances with $|r\rangle$ and $|f\rangle$ respectively, where $\Omega_r/2\pi = 20$ MHz, $\Omega_f/2\pi = 10$ MHz, $\delta_r/2\pi = (17 + \Delta)$ MHz, $\delta_f/2\pi = (15 + \Delta)$ MHz. The solid black line is a fit to data. **D** Interaction crossover in the 4-level system. Increasing the coupling to the non-interacting state $|f\rangle$, the effective potential amplitude, $V_0$ (blue curve), flips sign while the sign of the effective mass of the photons, $m_{EE}$ (red curve), remains the same. Inset: Dispersion relation (momentum $k$ vs $\Delta$) for **C** at $\Delta/2\pi = -16$ MHz (white dot) showing the two dark branches $|d_0\rangle$ and $|d_1\rangle$ (black lines). This dispersion is strongly modified with respect to the conventional 3-level EIT with $|r\rangle$ (blue curve) due to coupling to the non-interacting state $|f\rangle$. Dynamics due to propagation and Rydberg interactions further mix $|d_0\rangle$ and $|d_1\rangle$ which allows us to control the effective potential amplitude, $V_0$, and effective mass, $m_{EE}$, of photons.

Our experiments utilize a dense gas of cold $^{87}$Rb atoms in a crossed optical dipole trap (Fig.1A). The atoms are optically pumped into the hyperfine state, $F$, and magnetic sublevel, $m_F$, $|g\rangle = |5S_{1/2}, F = 1, m_F = 1\rangle$ of the electronic ground state (Fig.1B). A weak probe field at 780 nm off-resonantly couples the state $|g\rangle$ to the electronic excited state $|p\rangle = |5P_{3/2}, F = 2, m_F = 2\rangle$ off-detuned by $\Delta$. The $|p\rangle$ state is coupled to two metastable states: a Rydberg state $|r\rangle = |73S_{1/2}, m_J = 1/2\rangle$ by a control beam at 479 nm, resulting in strong van der Waals interactions; and a non-interacting ground state hyperfine sublevel $|f\rangle = |5S_{1/2}, F = 2, m_F = 2\rangle$ by another control beam at 780 nm, off-detuned relative to the two-photon frequency by $\delta_r$ and $\delta_f$ respectively. This coupling scheme results in the double-peaked transmission spectrum shown in Fig.1C.

To characterize quantum nonlinear effects, photon-photon correlations function are measured. Because dispersion outside of the atomic medium is negligible, any amplitude and phase features formed inside the nonlinear medium are preserved outside and can be detected in the form of photon number and phase correlations. We split the transmitted probe beam equally into three paths (Fig.1A). This allows us to measure the two- and three-photon correlation functions $g^{(2)}(\tau)$ or $g^{(3)}(\tau_{21}, \tau_{31})$ as needed. $\tau$ is the time separation between two photons and $\tau_{21}$ and $\tau_{31}$ are the time separations between the pairs of photons. For the phase correlations, we can additionally mix a frequency-shifted probe into one of the paths to act as a local oscillator (LO). We perform a heterodyne measurement to extract the conditional two-photon phase $\phi^{(2)}(\tau)$, which is the phase of a photon conditioned on detecting another photon at time $\tau$ away. (See Supplementary Information for details.)

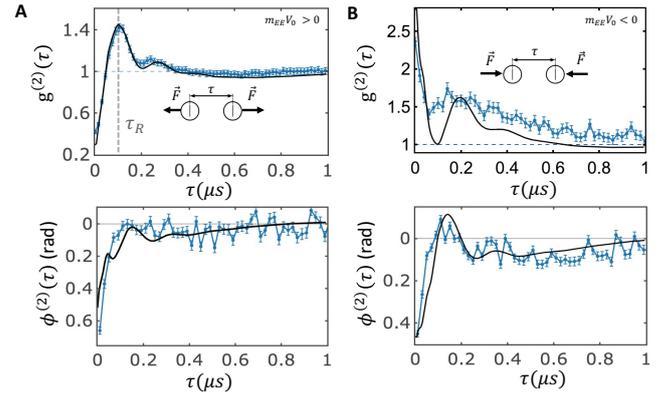

Figure 2: **Correlation functions showing repulsion and attraction.** Figures show the two-photon correlation function $g^{(2)}(\tau)$ and two-photon phase $\phi^{(2)}(\tau)$, where $\tau$ is the time separation between the two photons. **A**. Photon repulsion - $g^{(2)}(0) = 0.42 \pm 0.02$ is antibunched and peaks at later time $\tau_R = (0.10 \pm 0.02)$ $\mu$s with a value of $1.42 \pm 0.04$. **B**. Photon attraction - $g^{(2)}(0) = 2.6 \pm 0.08$ is bunched. The two photon phase is negative at $\tau = 0$ for both with values $\phi^{(2)}(0) = (-0.66 \pm 0.02)$ rad and $(-0.52 \pm 0.02)$ rad for the repulsion and attraction data respectively. Both data are taken at the parameters $\Omega_r/2\pi = 20$ MHz, $\Omega_f/2\pi = 10$ MHz, $\delta_r/2\pi = 1.1$ MHz, $\delta_f/2\pi = -1.1$ MHz and with $\Delta/2\pi = -10$ MHz for repulsion and $\Delta/2\pi = 10$ MHz for attraction. The black solid is a theory simulation of the system by propagating the two-particle optical Bloch equations. Simulations use experimental parameters, but $\Omega_f/2\pi$ is increased to 12 MHz to fit the data. Error bars in figure indicate one standard deviation (s.d.)

Fig.2 shows measurements of the two-photon correlation function $g^{(2)}$ for different parameters corresponding to regimes of attractive or repulsive interactions. Fig.2A presents evidence of photon repulsion in the form of an antibunching feature with a value $g^{(2)}(0) = 0.42 \pm 0.02$. The likelihood of finding two photons is suppressed at $\tau = 0$, but then peaks at a finite temporal separation defined as $\tau_R$, indicating that the photons have been pushed away from each other (see insets in Fig.2). Indeed we observe that $g^{(2)}$ exceeds unity at later times with a peak value $1.42 \pm 0.04$ at a temporal separation $\tau_R = (0.10 \pm 0.02)$ $\mu$s. Here the one-photon detuning $|\Delta|/2\pi = 10$ MHz is larger than the the decay rate of the state $|e\rangle$, $\Gamma/2\pi = 6.1$ MHz, ensures that the dissipative nonlinear effects are too small to affect the observed features of $g^{(2)}$. At different detuning parameters, Fig.2B demonstrates photon attraction where we measure $g^{(2)}(0)$ to be bunched with a value $g^{(2)}(0) = 2.6 \pm 0.08$, indicating enhanced likelihood to find two photons at the same position. In addition to $g^{(2)}$, the two-photon phase $\phi^{(2)}$ is measured to verify the coherent nature of the interactions. For both cases, we observe a negative conditional two-photon phase at zero time sep-aration $\tau = 0$, with values $\phi^{(2)}(\tau = 0) = (-0.66 \pm 0.02)$ rad for the repulsive case and $(-0.52\pm0.02)$ rad for the attractive case. This non-zero value of the conditional two-photon phase provides further evidence that the observed correlation features are due to coherent interactions, and not from nonlinear dissipation. The physical origin of the sign of the interaction is explained with an effective theoretical description presented below. Solid lines in Fig. 2 show the results of a full theoretical model (see SI), in which we numerically solve the set of propagation equations for the light field and atomic coherences. These simulations are in good agreement with our experimental results.

To provide more intuitive insight into the observed behavior of $g^{(2)}$ and $\phi^{(2)}$, we derive an effective theoretical description. Our scheme features two propagating dark states $|d_0\rangle$ and $|d_1\rangle$. Since they are nearly degenerate they are coupled to each other [26]. Thus we use a two-component wavefunction to describe propagation dynamics in the system, $\boldsymbol{\psi}(Z, z) \equiv \begin{pmatrix} \psi_1(Z,z) \\ \psi_2(Z,z) \end{pmatrix}$, where $Z$ is the mean position of the two excitations, and $z$ is their spatial separation. The basis of $\boldsymbol{\psi}(Z, z)$ is chosen such that the effective mass matrix $\overset{\leftrightarrow}{M}$ is diagonal, which allows one to interpret $\boldsymbol{\psi}(Z, z)$ as propagating massive particles. Furthermore, $\psi_1(Z,z)$ is proportional to the two-photon field amplitude $EE(Z,z)$ which allows us to directly compare the predictions to the observed photon dynamics (see SI). The dynamics of $\boldsymbol{\psi}(Z, z)$ are shown to be governed by a two-component Schrödinger-like equation:

$$iv_{\text{avg}}\partial_Z \boldsymbol{\psi}(Z,z) = -\frac{1}{2}\overset{\leftrightarrow}{M}^{-1} \partial_z^2 \boldsymbol{\psi}(Z,z) + (\overset{\leftrightarrow}{E}_0 + \overset{\leftrightarrow}{E}_v \tilde{V}(z))\boldsymbol{\psi}(Z,z), \quad (1)$$

where $v_{\text{avg}}$ is the dark-states' average group velocity $(v_{d0}+v_{d1})/2$, $\overset{\leftrightarrow}{E}_0$ is the non-interacting energy matrix, $\overset{\leftrightarrow}{E}_v$ is the interaction matrix, and $\tilde{V}(z)$ is the effective potential. Coupling between $\psi_1(Z,z)$ and $\psi_2(Z,z)$ manifests as off-diagonal terms in $\overset{\leftrightarrow}{E}_0$ and $\overset{\leftrightarrow}{E}_v$, the latter of which arises from Rydberg interactions. The van der Waals interaction between Rydberg states $V(z) = C_6/z^6$ is renormalized to yield the step-like effective potential with height $V_0$ given by $\tilde{V}(z) = \frac{V(z)}{1+V(z)/V_0}$ [17]. When the interactions shifts the Rydberg state away from resonance by an EIT linewidth, it eliminates transparency and saturates the effective potential. The diagonal terms of $\overset{\leftrightarrow}{E}_v$ are always positive, allowing us to adjust the sign of the effective potential via $V_0$. The effective mass for $\psi_1 \propto EE$ is equal to the reduced mass $m_{\text{EE}} = m_{d0}m_{d1}/(m_{d0}+m_{d1})$ of the dark-state excitations.

The observed repulsion and attraction in Figure 2 can be understood from equation (1) by considering the signs of $V_0$ and $m_{\text{EE}}$ as one would in a standard Schrodinger equation. In the case of repulsion we have the product $m_{\text{EE}}V_0 > 0$ and in the case of attraction we have $m_{\text{EE}}V_0 < 0$. However, the coupling between $\psi_1$ and $\psi_2$ in this double EIT-scheme offers richer dynamics compared to a conventional EIT scheme, where only one dark-state branch exists. The sign of the effective potential and the effective mass for $\psi_1$ are good qualitative indicators of the interactions, but the coupling to $\psi_2$ is crucial to describe the full dynamics. To see this, we scan the two-photon detuning $\delta_r$, with $\delta_f = -\delta_r$, at fixed one-photon detuning $\Delta/2\pi = -16$MHz (Fig.3). Within the scanned range, both $m_{\text{EE}}$ and $V_0$ remain positive, and therefore we expect repulsion between photons. Indeed we observe zero-time antibunching $g^{(2)}(0)$ and delayed-time positive correlations $g^{(2)}(\tau_R) > 1$ (Fig. 3A). However, the zero-time phase $\phi^{(2)}(0)$ switches from negative to positive as $\delta_r$ increases (Fig. 3B). This switch is captured by the full theoretical model and also the effective model (see SI). Such a sign flip of $\phi^{(2)}(0)$ with the signs of $V_0$ and $m_{\text{EE}}$ unchanged is absent in a one-component Schrodinger-like description [14].

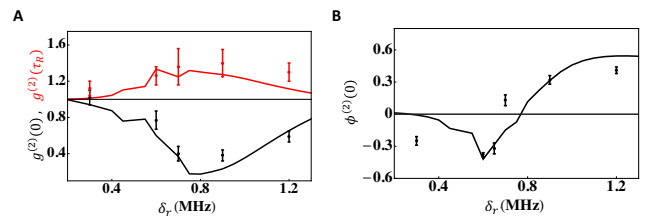

Figure 3: **Repulsive interactions and phase. A** Repulsive interactions as indicated by antibunching at $g^2(0) < 1$ and peak $g^2(\tau_R) > 1$. Results from full theoretical simulations are shown as solid black lines. Experimental parameters: $\Omega_r/2\pi = 23$ MHz, $\Omega_f/2\pi = 10$ MHz, $\Delta/2\pi = -16$ MHz, and with $\delta_f = -\delta_r$. $\Omega_f$ is increased to 11 MHz in theoretical simulation to fit the data and $\delta_f$ is corrected for a stark shift arising from imperfect polarization (see SI for details). For these parameters, we have $m_{\text{EE}}$ and $V_0$ remain positive implying repulsion as discussed in the main text. **B** Zero-time phase correlation $\phi^2(0)$ changes from negative to positive even when the sign of $m_{\text{EE}}$ and $V_0$ don't change, illustrating the need for a two-component wavefunction. Error bars in figure indicate one s.d.

$N$-particle correlations can serve as a powerful tool to characterize emerging self-ordered phases of matter [27]. In our system, we further investigate the quantum dynamics of repulsively interacting photons by measuring the third-order correlation function $g^{(3)}(\tau_{21}, \tau_{31})$. The experimentally measured $g^{(3)}$ for the same parameters used to measure $g^{(2)}$ in Fig.2A is shown in Fig.4A. We show $g^{(3)}$ as the 2D correlations of detecting three photons arriving with time separations $\tau_{21}$ and $\tau_{31}$. Just as in the case of the $g^{(2)}$ measurement, when two photons are detected together ($\tau_{21} = 0$ or $\tau_{31} = 0$ or $\tau_{21} - \tau_{31} = 0$ corresponding to the vertical, horizontal, and diagonal lines), we observe repulsion features of anti-bunching

(red) followed by positive correlations (blue) at a separation of $\tau_R$.

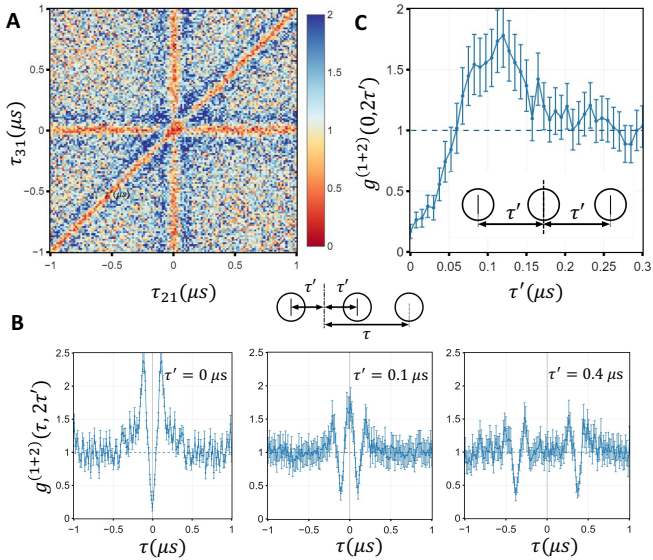

Figure 4: **Three-photon correlations and onset of crystal-like behaviour.** **A** 2D projection of the three-photon correlation function $g^{(3)}(\tau_{21}, \tau_{31})$ for the same parameters as in Fig.2A. **B** Likelihood of detecting a photon at separation of $\tau$ from the midpoint of two other photons detected with separation $2\tau'$: $g^{(1+2)}(\tau, 2\tau') = g^{(3)}(\tau+\tau', 2\tau')/g^{(2)}(2\tau')$. Probabilities are shown for $2\tau' = 0.00(1), 0.20(1), 0.80(1)$ $\mu s$ indicating 3-photon wavefunction interference. See text for discussion. **C** Likelihood of three photons detected at equal intervals to be $\tau'$, which is maximum near the characteristic timescale $\tau_R = (0.10 \pm 0.02)$ $\mu s$. Error bars in figure indicate one s.d.

To investigate interference effects between three-photons, we analyze the situation where two photons are detected with a time separation $2\tau'$, and plot in Fig. 4B the likelihood to detect the third photon detected at time $\tau$ from their midpoint. We define this normalized correlation as $g^{(1+2)}(\tau, 2\tau')$, corresponding to $g^{(3)}(\tau'+\tau, 2\tau')$ (averaged over all permutations) normalized by the two-photon correlation $g^{(2)}(2\tau')$. Fig.4B shows $g^{(1+2)}$ for a few different time separations $\tau'$. For long time separations ($\tau' > 2\tau_R$), the third photon experiences repulsion near either photon of the pair independently. For very short-time separations ($\tau' \sim 0$), the third photon experiences a much stronger repulsion from the pair than what we observe in $g^{(2)}$, as evidenced by the enhancement of both the anti-bunching ($0.18 \pm 0.04$) and bunching features ($2.37 \pm 0.16$) of $g^{(1+2)}$. Compared to the peak value of $g^{(2)}$ ($1.42 \pm 0.04$), the deviation from 1 is more than twice larger. This indicates that the 3-photon system is governed by 3-photon wavefunction interference, rather than a statistical mixture of two-photon repulsive processes. As the pair separates, the third photon gets pushed from either side, which increases its likelihood of being localized in between the two other photons. Fig.4C plots the likelihood of three photons detected with an equal interval separated by $\tau'$ by scanning $g^{(1+2)}(0, 2\tau')$. As $\tau'$ reaches the time separation $\tau_R$, the third photon is most likely to be in the middle of the other two photons with a maximum value of $g^{(1+2)}(0, 2\tau') = 1.76 \pm 0.2$, near $2\tau_R$ within experimental uncertainty. This indicates the presence of coherent 3-photon wavefunction interference, and hints at the onset of crystallization of photons mediated through repulsive interactions.

The ability to tune photon-photon interactions across attractive and repulsive regimes provides an opportunity for extensive control of strongly interacting photons in both 1D and higher dimensions. This opens avenues to studying exotic phases of matter [28], including self-organization in open quantum systems [3], as well as photonic quantum materials that cannot be realized with conventional quantum systems. Many-body states like photon crystals can enable new applications in quantum metrology and quantum communication [11, 12]. The extension to higher dimensions can be accomplished by carefully shaping the probing laser using spatial-light modulators [29]. By engineering the effective mass and interactions along different dimensions, it might be possible to sustain photonic Efimov states when using all-attractive forces or study topological physics with photons [30].

The authors acknowledge helpful conversations with Thomas Pohl and Callum Murray. The authors also acknowledge help with control electronics from Zhenpu Zhang.

Author contributions: The experiment and analysis were carried out by S.H.C, A.V.V, W.X, and B.J. Theoretical modelling was done by L.Z and A.V.V. All work was supervised by M.D.L and V.V. All authors discussed the results and contributed to the manuscript.

# Supplementary information to: Repulsive photons in a quantum nonlinear medium

Sergio H. Cantu,[1,][*] Aditya V. Venkatramani,[2,][*] Wenchao Xu,[1] Leo Zhou,[2] Brana Jelenković,[3] Mikhail D. Lukin,[2] and Vladan Vuletić[1]

[1]*Department of Physics and Research Laboratory of Electronics,*
*Massachusetts Institute of Technology, Cambridge, Massachusetts 02139, USA*
[2]*Department of Physics, Harvard University, Cambridge, Massachusetts 02138, USA*
[3]*Photonics Center, Institute of Physics, University of Belgrade, Serbia*

### Contents

## I. METHODS

### A. Atom preparation

[87]Rb atoms are cooled in a 3D magneto-optical trap (MOT) and loaded into a far-off detuned 1064 nm crossed dipole trap with an opening angle of $32°$. This results in a cigar-shaped atomic cloud with dimensions root-mean-squared (RMS) axial width of 8 $\mu m$ and radial width of 32 $\mu m$ with a optical depth ($OD$) of $\sim 30$. The cloud is cooled to 50 $\mu K$ using polarization gradient cooling to reduce Doppler broadening of atomic transitions.

We apply a magnetic field of 15.5 Gauss along the direction of propagation of the probe to set our quantization axis. The magnitude is chosen to separate the magnetic Zeeman levels sufficiently to minimize effects from other states. The atoms are optically pumped (Fig.1A) into the hyperfine ($F$) and Zeeman ($m_F$) sublevel $|g\rangle = |5S_{1/2}, F = 1, m_F = 1\rangle$. A weak probe field ($\approx 1$ ph $\mu sec^{-1}$) which is at 780 nm and $\sigma_+$-polarized, addresses $|g\rangle$ to the intermediate state $|p\rangle = |5P_{3/2}, F = 2, m_F = 2\rangle$. The probe is coupled to the Rydberg state $|r\rangle = |73S_{1/2}, m_J = 1/2\rangle$ by a counter $\sigma_-$ polarized propagating control field at 479 nm. The probe field is also coupled to a non-interacting hyperfine ground state $|f\rangle = |5S_{1/2}, F = 2, m_F = 2\rangle$, by a $\pi$-polarized control field at 780 nm applied perpendicularly. At single-photon detuning of $\Delta/2\pi = -16$ MHz, and our quantization field of 15.5 Gauss, the control laser coupling $|f\rangle$ and $|e\rangle$ also couples $|f\rangle$ and $|5P_{3/2}, F = 2, m_F = 1\rangle$ from residual $\sigma_-$-polarization of $\approx 1\%$ compared to the expected $\pi$-polarization. $\delta_f$ in the numerics is corrected for the stark shift arising from this spurious coupling, which shifts $\delta_f$ by $\approx 300$ KHz for parameters in Fig.3 of the main text.

The Rydberg state $|r\rangle$ strongly interacts with a Van-der Waals interaction $V(z) = C_6/z^6$, where $C_6/\hbar = 2\pi \times 1.8$ THz $\mu m^6$. The probe beam is focused to a waist of $\omega \sim 4.5$ $\mu m$, smaller than the blockade radius of $\sim 10$ $\mu m$, resulting in an effective 1D system for the propagation of the probe polariton. We send a probe pulse for 6 $\mu$s repeating every 40 $\mu$s. The dipole trap is turned off during probing to prevent anti-trapping atoms in the Rydberg state and non-homogeneous AC stark shifts of the states. We repeat this 1500 times every 1.5 seconds before we have to reload and cool atoms into the dipole trap again.

---

[*] These authors contributed equally to this work

## B. Correlation measurements

To study the correlations between photons after they pass through the atomic gas (Fig.1A), we split the beam into three paths. This allows us to measure two and three photon correlation function $g^2$ and $g^3$. Suppose detectors 1,2,3 detect $n_1, n_2, n_3$ photons at times $t_1, t_2, t_3$, then $g^2(t_2 - t_1) = \frac{\langle n_1(t_1)n_2(t_2)\rangle}{\langle n_1(t_1)\rangle\langle n_2(t_2)\rangle}$, where $\langle\rangle$ denotes averages over multiple experimental repeats. $g^2$ can similarly be defined over all combinations of pairs of detectors. $g^3(t_2 - t_1, t_3 - t_1) = \frac{\langle n_1(t_1)n_2(t_2)n_3(t_3)\rangle}{\langle n_1(t_1)\rangle\langle n_2(t_2)\rangle\langle n_3(t_3)\rangle}$.

To measure the conditional phase of the photons, we send a local oscillator (LO) co-propagating alongside the the probe. The LO is detuned 80 MHz away from the probe and propagates with orthogonal polarization to suppress photon scattering from the atomic cloud. The LO is then mixed into one of the detectors ($d_1$) using a 8:92 pellicle beamsplitter. We perform a heterodyne measurement to obtain the phase of the probe beam as a function of time $t_1$. This phase can be conditioned on detecting a photon on either one of the other detectors at time $t_2$ to give us the conditional phase $\phi^2(t_2 - t_1)$.

These correlation functions can be related to the two-photon wave function. Let us denote $E(z)$ as the probability amplitude of having a photon at position $z$. This can be extended to two photons by the probability amplitude $EE(z_2 - z_1)$, which would correspond to having two photons at positions $z_1$ and $z_2$. Then $g^2(t_2 - t_1) = \left|\frac{EE(c(t_2-t_1))}{E(ct_1)E(ct_2)}\right|^2$, and $\phi^2(t_2 - t_1) = \text{Arg}\left(\frac{EE(c(t_2-t_1))}{E(ct_1)E(ct_2)}\right)$. Measuring $g^2$ and $\phi^2$ directly gives us information about the two photon amplitude. This definition can be analogously extended to $g^3$ as well.

## II. A TWO-COMPONENT EFFECTIVE EQUATION GOVERNING POLARITON DYNAMICS

In this section, we derive the effective theoretical description of polariton dynamics that we experimentally observe. We start with the two-body equations of motion that has 16 components, and perform a series of approximations and simplication that culminates in the two-component Schrödinger-like equation (1) in the main text.

As the experiments are conducted in the regime where the waist of the probe beam is much smaller than the Rydberg blockade radius in the atomic medium, we assume the dynamics of quasi-particle excitations are confined to one dimension to good approximation. In the context of the 4-level scheme shown in Figure 1B, let us denote $\hat{\mathcal{E}}^\dagger(z), \hat{\mathcal{P}}^\dagger(z), \hat{\mathcal{R}}^\dagger(z), \hat{\mathcal{F}}^\dagger(z)$ as the creation operator of a photon, an intermediate-state excitation $|p\rangle$, a Rydberg excitation $|r\rangle$, and an excitation in the non-interacting ground state $|f\rangle$, respectively, at position $z$. These operators satisfy the bosonic commutation relation $[\hat{\mathcal{E}}(z), \hat{\mathcal{E}}^\dagger(z')] = [\hat{\mathcal{P}}(z), \hat{\mathcal{P}}^\dagger(z')] = [\hat{\mathcal{R}}(z), \hat{\mathcal{R}}^\dagger(z')] = [\hat{\mathcal{F}}(z), \hat{\mathcal{F}}^\dagger(z')] = \delta(z - z')$.

Under the scheme shown in Fig. 1B, the Hamiltonian governing the system within the atomic medium is

$$\mathcal{H} = \mathcal{H}_0 + \mathcal{H}_{\text{int}}, \quad \text{where} \quad \mathcal{H}_0 = \int dz \begin{pmatrix}\hat{\mathcal{E}}\\\hat{\mathcal{P}}\\\hat{\mathcal{R}}\\\hat{\mathcal{F}}\end{pmatrix}^\dagger \begin{pmatrix} -ic\partial_z & g/2 & 0 & 0 \\ g/2 & -\Delta & \Omega_r/2 & \Omega_f/2 \\ 0 & \Omega_r/2 & -\delta_r & 0 \\ 0 & \Omega_f/2 & 0 & -\delta_f \end{pmatrix} \begin{pmatrix}\hat{\mathcal{E}}\\\hat{\mathcal{P}}\\\hat{\mathcal{R}}\\\hat{\mathcal{F}}\end{pmatrix} \quad (S1)$$

$$\text{and} \quad \mathcal{H}_{\text{int}} = \frac{1}{2}\iint dz\, dz'\, V(z-z')\hat{\mathcal{R}}^\dagger(z)\hat{\mathcal{R}}^\dagger(z')\hat{\mathcal{R}}(z')\hat{\mathcal{R}}(z) \quad (S2)$$

where $g$ is the collective photon-atom coupling determined by the atomic density resonant atomic cross section. In our experimental regime of high optical depth OD = 30, $g$ is larger than the other parameters in the Hamiltonian $\mathcal{H}_0$ by an of magnitude.

In the Heisenberg picture, the particle operators obey the following Heisenberg equations of motion:

$$i\partial_t \hat{\mathcal{E}} = -ic\partial_z \hat{\mathcal{E}} + \frac{g}{2}\hat{\mathcal{P}} \quad (S3)$$

$$i\partial_t \hat{\mathcal{P}} = -\Delta\hat{\mathcal{P}} + \frac{g}{2}\hat{\mathcal{E}} + \frac{\Omega_r}{2}\hat{\mathcal{R}} + \frac{\Omega_f}{2}\hat{\mathcal{F}} \quad (S4)$$

$$i\partial_t \hat{\mathcal{R}} = -\delta_c \hat{\mathcal{R}} + \frac{\Omega_r}{2}\hat{\mathcal{P}} + \int dz'\, V(z-z')\hat{\mathcal{R}}^\dagger(z')\hat{\mathcal{R}}(z')\hat{\mathcal{R}}(z) \quad (S5)$$

$$i\partial_t \hat{\mathcal{F}} = -\delta_d \hat{\mathcal{F}} + \frac{\Omega_f}{2}\hat{\mathcal{P}} \quad (S6)$$

We now make the approximation of adiabatically eliminating the intermediate-state excitation $\hat{\mathcal{P}}$ by setting its time-derivative to zero. While this is typically justified when $|\Delta| \gg \Omega_r, \Omega_f$, we have found this to be a good approximation even in the $|\Delta| \sim \Omega_r, \Omega_f$ regime, as verified by comparing numerical simulations of two-particle problem with the full set of equations and that with $\hat{\mathcal{P}}$ adiabatically eliminated. We obtain:

$$i\partial_t \hat{\mathcal{E}} = -ic\partial_z \hat{\mathcal{E}} + \frac{g^2}{4\Delta}\hat{\mathcal{E}} + \frac{g\Omega_r}{4\Delta}\hat{\mathcal{R}} + \frac{g\Omega_f}{4\Delta}\hat{\mathcal{F}} \tag{S7}$$

$$i\partial_t \hat{\mathcal{R}} = -\delta_r \hat{\mathcal{R}} + \frac{g\Omega_r}{4\Delta}\hat{\mathcal{E}} + \frac{\Omega_r^2}{4\Delta}\hat{\mathcal{R}} + \frac{\Omega_r \Omega_f}{4\Delta}\hat{\mathcal{F}} + \int dz' V(z-z')\hat{\mathcal{R}}^\dagger(z')\hat{\mathcal{R}}(z')\hat{\mathcal{R}}(z) \tag{S8}$$

$$i\partial_t \hat{\mathcal{F}} = -\delta_f \hat{\mathcal{F}} + \frac{g\Omega_f}{4\Delta}\hat{\mathcal{E}} + \frac{\Omega_f^2}{4\Delta}\hat{\mathcal{F}} + \frac{\Omega_r \Omega_f}{4\Delta}\hat{\mathcal{R}} \tag{S9}$$

This indicates that the effective Hamiltonian under this approximation is $\mathcal{H}' = \mathcal{H}_0' + \mathcal{H}_{\text{int}}$

$$\mathcal{H}_0' = \mathcal{H}_0 = \int dz \begin{pmatrix} \hat{\mathcal{E}} \\ \hat{\mathcal{R}} \\ \hat{\mathcal{F}} \end{pmatrix}^\dagger \begin{pmatrix} -ic\partial_z + g^2/4\Delta & g\Omega_r/4\Delta & g\Omega_f/4\Delta \\ g\Omega_r/4\Delta & -\delta_r + \Omega_r^2/4\Delta & \Omega_r \Omega_f/4\Delta \\ g\Omega_f/2 & \Omega_r \Omega_f/4\Delta & -\delta_f + \Omega_f^2/4\Delta \end{pmatrix} \begin{pmatrix} \hat{\mathcal{E}} \\ \hat{\mathcal{R}} \\ \hat{\mathcal{F}} \end{pmatrix} \tag{S10}$$

which we will use for the remainder of the derivation.

### A. Single-particle dynamics

Let us first look at a single particle case, for which the wavefunction takes the form

$$|\psi\rangle = \int dz \sum_{A \in \{E,R,F\}} A(z,t) \hat{\mathcal{A}}^\dagger(z) |0\rangle \tag{S11}$$

We perform the following transformation on the coefficients:

$$A(z,t) = \sum_{k,\omega} A(k,\omega) e^{i(k+k_0)z - i\omega t} \quad \text{where} \quad k_0 = \frac{-g^2 \delta_r \delta_f}{c\Gamma_1} \quad \text{and} \quad \Gamma_1 = 4\delta_r \delta_f \left(\Delta - \frac{\Omega_r^2}{4\delta_r} - \frac{\Omega_f^2}{4\delta_f}\right) \tag{S12}$$

The added momentum shift of $k_0$ makes it so that in the new momentum-frequency $(k,\omega)$ basis, the zero-frequency ($\omega = 0$) eigenstate is at zero momentum ($k = 0$). For the single-particle case, we get the following equations of motion for the coefficients:

$$\omega E = \left(\frac{g^2}{4\Delta} + k_0 c + kc\right) E + \frac{g\Omega_r}{4\Delta} R + \frac{g\Omega_f}{4\Delta} F \tag{S13}$$

$$\omega R = -\delta_r R + \frac{g\Omega_r}{4\Delta} E + \frac{\Omega_r^2}{4\Delta} R + \frac{\Omega_r \Omega_f}{4\Delta} F \tag{S14}$$

$$\omega F = -\delta_f F + \frac{g\Omega_f}{4\Delta} E + \frac{\Omega_f^2}{4\Delta} F + \frac{\Omega_r \Omega_f}{4\Delta} R \tag{S15}$$

Solving $k$ as a function of $\omega$ will give us the single-particle dispersion relation for our system (shown in Fig. 1D):

$$k(\omega) = \frac{1}{c} \frac{(4\Delta\omega - g^2)(\delta_f + \omega)(\delta_r + \omega) - \omega\Omega_f^2(\delta_r + \omega) - \omega\Omega_r^2(\delta_f + \omega)}{4\Delta(\delta_f + \omega)(\delta_r + \omega) - \Omega_f^2(\delta_r + \omega) - \Omega_r^2(\delta_f + \omega)} - k_0 \tag{S16}$$

where, as mentioned earlier, $k_0 = -g^2 \delta_r \delta_f / c\Gamma_1$ is chosen so that $k(0) = 0$.

We can solve the above equation for $\omega$ to obtain eigenvalues $\omega = \omega_i(k)$ as a function of $k$, where $i = 1, 2, 3$ corresponds to three allowed values of $\omega$. The constant $k_0$ was chosen so that at $k = 0$, one of the eigenvalue $\omega_i(0) = 0$. We then identify the eigenstate with $\omega = 0$ as a dark state $|d_0\rangle$ with energy $E_{d_0} = 0$ at $k = 0$. In the limit of large collective atom-photon coupling $g$, the other two values of $\omega_i(k = 0)$ are (to leading order in $g$)

$$E_{d_1} = -\frac{\delta_f^2 \Omega_r^2 + \delta_r^2 \Omega_f^2}{\delta_f \Omega_r^2 + \delta_r \Omega_f^2} + O(g^{-2}), \quad E_b = \frac{g^2}{4\Delta} + k_0 c + O(g^0), \tag{S17}$$

We note that $E_b \gg E_{d_1}$, and identify the corresponding eigenstates $|b\rangle$ as a bright state and $|d_1\rangle$ as another dark state.

These eigenstates can be smoothly continued as a function of $k$ to obtain eigenstates $|d_{0k}\rangle$, $|d_{1k}\rangle$, and $|b_k\rangle$. This gives rise to the momentum dependence of the dark and bright state energies, from which we can calculate their group velocities and effective masses (at zero-momentum) as defined by

$$v = \frac{\partial \omega}{\partial k}\bigg|_{k\to 0} \quad \text{and} \quad m^{-1} = \frac{\partial^2 \omega}{\partial k^2}\bigg|_{k\to 0} \tag{S18}$$

The expressions for the group velocities of the three states are written below (to leading order in $g$):

$$v_{d0} = \frac{c\Gamma_1^2}{g^2(\delta_f^2 \Omega_r^2 + \delta_r^2 \Omega_f^2)} + O(g^{-4}), \quad , v_{d1} = v_{d0}\alpha_x^2 + O(g^{-4}), \quad v_b = c + O(g^{-2}), \tag{S19}$$

where

$$\alpha_x = \frac{\Omega_r \Omega_f (\delta_r - \delta_f)}{\delta_f \Omega_r^2 + \delta_r \Omega_f^2}, \tag{S20}$$

Although the expressions for $E_i$ and $v_i$ are given approximately above, we can write expressions for the effective masses for the three states exactly in terms of $E_i$ and $v_i$ as:

$$m_{d0}^{-1} = 2\left(\frac{v_{d0}v_{d1}}{0 - E_{d1}} + \frac{v_{d0}v_b}{0 - E_b}\right), \quad m_{d1}^{-1} = 2\left(\frac{v_{d1}v_{d0}}{E_{d1} - 0} + \frac{v_{d1}v_b}{E_{d1} - E_b}\right), \quad m_b^{-1} = 2\left(\frac{v_b v_{d0}}{E_b - 0} + \frac{v_b v_{d1}}{E_b - E_{d1}}\right). \tag{S21}$$

### B. Two-particle dynamics

In general, the two-particle dynamics for 4-level systems is described by a 16-component system of differential equations. As we have adiabatically eliminated the intermediate-state excitation $\hat{\mathcal{P}}$, we are left with an effective 3-level system with a 9-component system of equations. At the end of the section, we will arrive at a 2-component effective theory written in Eq. (S53) that describes the physics of the 16-component system in steady state limit, with some approximations that we will elaborate on.

We begin by writing the two-particle wavefunction in the following form:

$$|\psi\rangle = \int dz_1 dz_2 e^{ik_0(z_1+z_2)} \left[\sum_{A<B} AB(z_1, z_2, t)\hat{\mathcal{A}}^\dagger(z_1)\hat{\mathcal{B}}^\dagger(z_2) + \frac{1}{2}\sum_A AA(z_1, z_2, t)\hat{\mathcal{A}}^\dagger(z_1)\hat{\mathcal{A}}^\dagger(z_2)\right]|0\rangle \tag{S22}$$

Analogous to the single-particle case, the factor of $e^{ik_0(z_1+z_2)}$ is introduced so that in the non-interacting limit ($V = 0$), there is a zero-energy eigenstate at $k_1 = k_2 = 0$. In addition, we may assume without loss of generality that $AA(z_1, z_2, t) = AA(z_2, z_1, t)$ is symmetric in the two spatial coordinates due to the canonical commutation relation of the bosonic operators. To write down the two-particle equations of motion, it is convenient to switch to the center of mass $Z$ and relative coordinates $z$:

$$Z = \frac{1}{2}(z_1 + z_2), \quad z = z_2 - z_1 \tag{S23}$$

$$\partial_Z = \partial_{z_1} + \partial_{z_2}, \quad \partial_z = \frac{1}{2}(\partial_{z_2} - \partial_{z_1}) \tag{S24}$$

It is also convenient to define $ER_\pm \equiv ER \pm RE$, $EF_\pm \equiv EF \pm FE$, and $RF_\pm \equiv RF \pm FR$. Note the identity

$$\partial_{z_1}AB \pm \partial_{z_2}BA = \frac{1}{2}\partial_Z AB_\pm - \partial_z AB_\mp. \tag{S25}$$

We then get the following two-particle equations of motion:

$$i\partial_t EE = \left[-ic\partial_Z + 2k_0 c + \frac{g^2}{2\Delta}\right]EE + \frac{g\Omega_r}{4\Delta}ER_+ + \frac{g\Omega_f}{4\Delta}EF_+ \tag{S26}$$

$$i\partial_t RR = \left[V(z) - 2\delta_r + \frac{2\Omega_r^2}{4\Delta}\right]RR + \frac{g\Omega_r}{4\Delta}ER_+ + \frac{\Omega_r\Omega_f}{4\Delta}RF_+ \tag{S27}$$

$$i\partial_t FF = \left[-2\delta_f + \frac{\Omega_f^2}{2\Delta}\right]FF + \frac{g\Omega_f}{4\Delta}EF_+ + \frac{\Omega_r\Omega_f}{4\Delta}RF_+ \tag{S28}$$

$$i\partial_t RF_+ = \left[\frac{\Omega_r^2 + \Omega_f^2}{4\Delta} - (\delta_r + \delta_f)\right] RF_+ + \frac{g\Omega_r}{4\Delta} EF_+ + \frac{g\Omega_f}{4\Delta} ER_+ + \frac{2\Omega_r \Omega_f}{4\Delta}(FF + RR) \quad \text{(S29)}$$

$$i\partial_t RF_- = \left[\frac{\Omega_r^2 + \Omega_f^2}{4\Delta} - (\delta_r + \delta_f)\right] RF_- + \frac{g\Omega_r}{4\Delta} EF_- - \frac{g\Omega_f}{4\Delta} ER_- \quad \text{(S30)}$$

$$i\partial_t ER_+ = \left[-i\frac{c}{2}\partial_Z + k_0 c - \delta_r + \frac{\Omega_r^2}{4\Delta} + \frac{g^2}{4\Delta}\right] ER_+ + ic\partial_z ER_- + \frac{g\Omega_r}{2\Delta}(RR+EE) + \frac{g\Omega_f}{4\Delta} RF_+ + \frac{\Omega_r \Omega_f}{4\Delta} EF_+ \quad \text{(S31)}$$

$$i\partial_t EF_+ = \left[-i\frac{c}{2}\partial_Z + k_0 c - \delta_f + \frac{\Omega_f^2}{4\Delta} + \frac{g^2}{4\Delta}\right] EF_+ + ic\partial_z EF_- + \frac{g\Omega_f}{2\Delta}(FF+EE) + \frac{\Omega_r \Omega_f}{4\Delta} ER_+ + \frac{g\Omega_r}{4\Delta} RF_+ \quad \text{(S32)}$$

$$i\partial_t ER_- = \left[-i\frac{c}{2}\partial_Z + k_0 c - \delta_r + \frac{\Omega_r^2}{4\Delta} + \frac{g^2}{4\Delta}\right] ER_- + ic\partial_z ER_+ - \frac{g\Omega_f}{4\Delta} RF_- + \frac{\Omega_r \Omega_f}{4\Delta} EF_- \quad \text{(S33)}$$

$$i\partial_t EF_- = \left[-i\frac{c}{2}\partial_Z + k_0 c - \delta_f + \frac{\Omega_f^2}{4\Delta} + \frac{g^2}{4\Delta}\right] EF_- + ic\partial_z EF_+ + \frac{g\Omega_r}{4\Delta} RF_- + \frac{\Omega_r \Omega_f}{4\Delta} ER_- \quad \text{(S34)}$$

*a. Solving for* $(EE, RR, FF, RF_+, RF_-)$ — We will first take the steady state limit by setting $\partial_t = 0$.

In the large $g$ limit, the "energy" term of $EE$, $g^2/2\Delta + 2k_0 c$, is large compared to the rest, which allows us to make the approximation that $\partial_Z EE = 0$, analogous to adiabatic elimination. We have verified the validity of this approximation by looking at numerical solutions of these differential equations with and without the $\partial_Z EE = 0$ assumption, and finding them to agree qualitatively.

With these simplifications, we can use Eqs. (S26)-(S30) to express $EE, RR, FF, RF_+, RF_-$ in terms of $ER_+, ER_-, EF_+, EF_-$. We can then reduce Eqs. (S31)-(S34) to

$$\frac{ic}{2}\partial_Z \begin{pmatrix} \psi_+ \\ \psi_- \end{pmatrix} - ic \begin{pmatrix} 0 & \mathbb{1} \\ \mathbb{1} & 0 \end{pmatrix} \partial_z \begin{pmatrix} \psi_+ \\ \psi_- \end{pmatrix} = \left[\begin{pmatrix} H_{0+} & 0 \\ 0 & H_{0-} \end{pmatrix} + \begin{pmatrix} H_{V+} & 0 \\ 0 & 0 \end{pmatrix} \tilde{V}(r)\right] \begin{pmatrix} \psi_+ \\ \psi_- \end{pmatrix}, \quad \text{(S35)}$$

where $\psi_+$, $\psi_-$ are

$$\psi_+ = \begin{pmatrix} ER_+(Z, z) \\ EF_+(Z, z) \end{pmatrix}, \qquad \psi_- = \begin{pmatrix} ER_-(Z, z) \\ EF_-(Z, z) \end{pmatrix}. \quad \text{(S36)}$$

For the non-interacting part, we have

$$H_{0+} = J_0 \delta_r^2 \delta_f^2 \begin{pmatrix} \Omega_f^2/\delta_f^2 & -(\Omega_r/\delta_r)(\Omega_f/\delta_f) \\ -(\Omega_r/\delta_r)(\Omega_f/\delta_f) & \Omega_r^2/\delta_r^2 \end{pmatrix}, \quad \text{(S37)}$$

where

$$J_0 \equiv \frac{g^2}{\Gamma_1 \Gamma_2} - \frac{1}{\delta_f \Omega_r^2 + \delta_r \Omega_f^2}, \quad \Gamma_2 = -4\Delta\left[\delta_r + \delta_f - \frac{\Omega_r^2 + \Omega_f^2}{4\Delta}\right], \quad \Gamma_1 = 4\delta_r \delta_f \left(\Delta - \frac{\Omega_r^2}{4\delta_r} - \frac{\Omega_f^2}{4\delta_f}\right). \quad \text{(S38)}$$

and

$$H_{0-} = \frac{1}{2}\left[\left(\frac{g^2}{\Gamma_2} + 1\right)\frac{\Omega_r^2 - \Omega_f^2}{4\Delta} + \delta_f - \delta_r\right]\sigma_z + \left(\frac{g^2}{\Gamma_2} + 1\right)\frac{\Omega_r \Omega_f}{8\Delta}\sigma_x + C_{0-}\mathbb{1} \quad \text{(S39)}$$

$$\text{where} \quad C_{0-} = \frac{g^2}{2}\left(\frac{1}{4\Delta} - \frac{\delta_r + \delta_f}{\Gamma_2}\right) + \frac{\Gamma_2}{8\Delta} + k_0 c. \quad \text{(S40)}$$

For the interacting part, we have

$$H_{V+} = \begin{pmatrix} \alpha_{1v} & \alpha_{2v} \\ \alpha_{2v} & \alpha_{5v} \end{pmatrix}, \quad \tilde{V}(z) = \frac{V(z)}{1 + V(z)/V_0} \quad \text{(S41)}$$

where

$$V_0 = \frac{2\Gamma_1 \Gamma_2}{4\Delta\Gamma_1 + (4\Delta\delta_f - \Omega_f^2)^2} \quad \text{(S42)}$$

and

$$\alpha_{1v} = \frac{g^2 \Omega_r^2 (\delta_f \Gamma_2 + \delta_r \Omega_f^2)^2}{2\Gamma_1^2 \Gamma_2^2}, \quad \alpha_{2v} = \frac{-g^2 \delta_f \Omega_r^3 \Omega_f (\delta_f \Gamma_2 + \delta_r \Omega_f^2)}{2\Gamma_1^2 \Gamma_2^2}, \quad \alpha_{5v} = \frac{g^2 \delta_f^2 \Omega_r^4 \Omega_f^2}{2\Gamma_1^2 \Gamma_2^2} \quad \text{(S43)}$$

b. *Eliminating $ER_-$ and $EF_-$ to get two-component theory* — In the large $g$ limit, using the parameters we usually work with, the diagonal terms of $H_{0-}$ are much larger than $H_{0+}$. This allows us to further make the approximation that $\partial_Z \psi_- = 0$, similar to adiabatic elimination. With this approximation, Eq. (S35) reduces to the Hermitian two-component equation:

$$\frac{ic}{2}\partial_Z \psi_+(Z,z) - c^2 H_{0-}^{-1} \partial_z^2 \psi_+(Z,z) = \left[H_{0+} + H_{V+}\tilde{V}(z)\right]\psi_+(Z,z), \tag{S44}$$

We are now very close to the final form of our two-component effective theory. To move further, we note that from Eq. (S26), under the approximation of setting $\partial_Z EE = 0$ and $\partial_t EE = 0$ as we have done earlier, we can write $EE$ as

$$EE = \frac{g/(4\Delta)}{-2(g^2/(4\Delta) + k_0 c)}(\Omega_r ER_+ + \Omega_f EF_+) \tag{S45}$$

Thus, we believe that it will be suggestive to transform the basis of $ER_+$ and $EF_+$ using the rotation matrix

$$U = \frac{g/(4\Delta)}{-2(g^2/(4\Delta) + k_0 c)}\begin{pmatrix} \Omega_r & \Omega_f \\ -\Omega_f & \Omega_r \end{pmatrix}, \tag{S46}$$

Using this transformation, we define

$$\boldsymbol{\psi}(Z,z) = U\psi_+(Z,z), \tag{S47}$$

giving

$$\boldsymbol{\psi}(Z,z) = \begin{pmatrix} EE(Z,z) \\ \frac{\Gamma_1}{2g(\delta_f \Omega_r^2 + \delta_r \Omega_f^2)}(-\Omega_f ER_+(Z,z) + \Omega_r EF_+(Z,z)) \end{pmatrix} \equiv \begin{pmatrix} \psi_1(Z,z) \\ \psi_2(Z,z) \end{pmatrix} \tag{S48}$$

Conveniently, we find that in the large $g$ limit, $U$ approximately diagonalizes $H_{0-}$ (see Eq. (S39)):

$$\overleftrightarrow{M}^{-1} = -2v_{\text{avg}}(UH_{0-}^{-1}U^{-1}) \approx \begin{pmatrix} \frac{1}{2m_{\text{EE}}} & 0 \\ 0 & \frac{1}{2m_2} \end{pmatrix} + O(g^{-6}), \tag{S49}$$

where

$$v_{\text{avg}} = \frac{v_{d0} + v_{d1}}{2}, \quad m_{\text{EE}} = \frac{m_{d0}m_{d1}}{m_{d0} + m_{d1}}, \quad m_2 = \frac{-g^4(\delta_f \Omega_r^2 + \delta_r \Omega_f^2)^2(\delta_f^2 \Omega_r^2 + \delta_r^2 \Omega_f^2)}{2c^2(\Omega_r^2 + \Omega_f^2)\Gamma_2 \Gamma_1^3}. \tag{S50}$$

We now describe the remaining expressions in our two-component equation of Eq. (S53) in the rotated basis:

$$\overleftrightarrow{E}_0 = \frac{2v_{\text{avg}}}{c}(UH_{0+}U^{-1}) \approx \frac{\Gamma_1}{\Gamma_2}\begin{pmatrix} \alpha_x^2 & -\alpha_x \\ -\alpha_x & 1 \end{pmatrix} + O(g^{-2}) \tag{S51}$$

$$\overleftrightarrow{E}_v = \frac{2v_{\text{avg}}}{c}(UH_{v+}U^{-1}) \approx \frac{\Omega_r^2}{2\Gamma_2^2(\delta_f \Omega_r^2 + \delta_r \Omega_f^2)^2}$$
$$\begin{pmatrix} \Omega_r^2(\Gamma_1 + 4\Delta\delta_f^2)^2 & -\Omega_r\Omega_f(\Gamma_1 + 4\Delta\delta_f^2)((\delta_r + \delta_f)(4\Delta\delta_f - \Omega_f^2) - 2\delta_f\Omega_r^2) \\ -\Omega_r\Omega_f(\Gamma_1 + 4\Delta\delta_f^2)((\delta_r + \delta_f)(4\Delta\delta_f - \Omega_f^2) - 2\delta_f\Omega_r^2) & \Omega_f^2((\delta_r + \delta_f)(4\Delta\delta_f - \Omega_f^2) - 2\delta_f\Omega_r^2)^2 \end{pmatrix}$$
$$+ O(g^{-2}) \tag{S52}$$

The effective two-particle dynamics in our system can be then described by the following two-component Schrödinger equation:

$$\boxed{iv_{\text{avg}}\partial_Z \boldsymbol{\psi}(Z,z) = -\overleftrightarrow{M}^{-1}\partial_z^2 \boldsymbol{\psi}(Z,z) + (\overleftrightarrow{E}_0 + \overleftrightarrow{E}_v \tilde{V}(z))\boldsymbol{\psi}(Z,z),} \tag{S53}$$

This equation holds when our approximations hold, which we summarize here:

- intermediate-state excitation $\hat{\mathcal{P}}$ can be adiabatically eliminated
- steady state limit by setting all time-derivatives to zero: $\partial_t = 0$
- $\partial_Z EE = \partial_Z ER_- = \partial_Z EF_- = 0$

## C. Comparing full theory with two-component effective theory

The simulations in the main text are carried out using the full two-particle equations of motion (S26)-(S34). The approximations used to derive the effective two component equation (S53) makes quantitative comparisons with experiment difficult, but as discussed in the main text equation (S53) can provide insight into why we obtain repulsion and attraction between photons. In order to compare the full set of two-particle equations and the effective theory, we compare the simulation results of both over several values of single-photon detuning $\Delta$ and two-photon detuning to the Rydberg state $\delta_r$. We identify regions of repulsion, attraction, positive phase, and negative phase show in Figure S1. The good agreement between the two simulations in identifying the right features implies that we can use equation (S53) as a predictor for the nature of interactions between two photons.

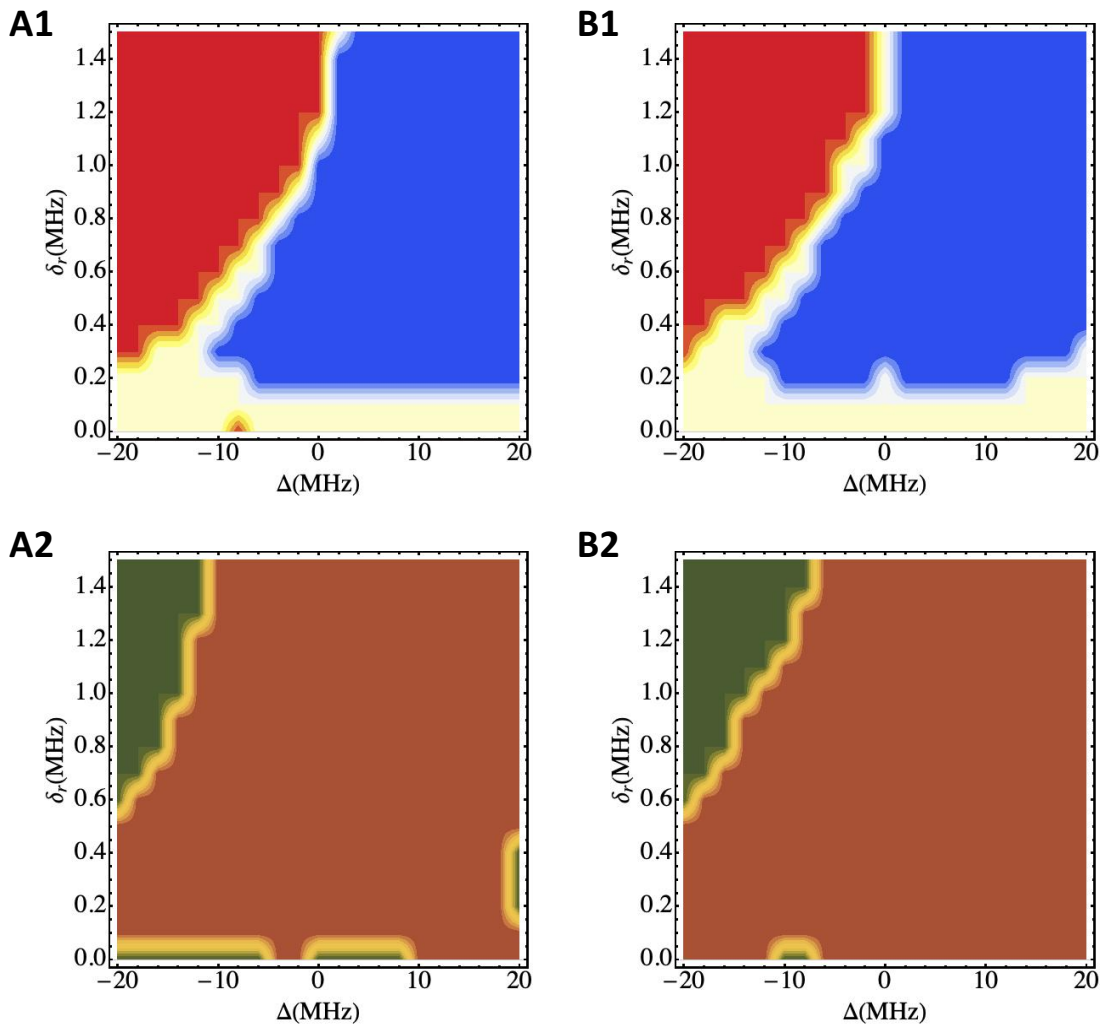

FIG. S1: **Comparing full numerics with effective theory. A1, B1** Comparing regions of effective interaction - red corresponds to repulsion, blue corresponds to attraction, and yellow corresponds to dissipation/ no correlation. **A1** is the result from the full numerics and **B1** is the result of simulating the effective theory. Repulsion is characterized by antibunching $g^2(0) < 0.95$ followed by bunching at a later time $g^2(\tau_R) > 1.05$. Attraction is characterized by the bunching $g^2(0) > 1.05$ and all other cases are characterized by dissipation or have no correlation. **A2, B2** Comparing sign of two photon phase - green is positive phase $\phi^2(0)$ and brown is negative phase $\phi^2(0)$. **A2** is the result from the full numerics and **B2** is the result of simulating the effective theory. Both simulations are carried out at parameters - $\Omega_r = 20$ MHz and $\Omega_f = 12$ MHz, $\delta_f = -\delta_r$.